\documentclass[aps,prb,reprint,amsmath,amssymb,superscriptaddress,longbibliography]{revtex4-2}

\usepackage{graphicx} 
\usepackage{dcolumn} 
\usepackage{bm} 
\usepackage{braket}

\usepackage{color}

\begin{document}

\title{Shift current conductivity in monolayer SnS: a tight-binding analysis}

\author{Yuki Kusunoki}
    \affiliation{
                Department of Nanotechnology for Sustainable Energy,
                School of Science and Technology, Kwansei Gakuin University,
                Gakuen-Uegahara 1, Sanda, 669-1330, Japan
        }%
\author{Tomoaki Kameda}
    \affiliation{
                Department of Nanotechnology for Sustainable Energy,
                School of Science and Technology, Kwansei Gakuin University,
                Gakuen-Uegahara 1, Sanda, 669-1330, Japan
        }%
\author{Katsunori Wakabayashi}%
   \affiliation{
                Department of Nanotechnology for Sustainable Energy,
                School of Science and Technology, Kwansei Gakuin University,
                Gakuen-Uegahara 1, Sanda, 669-1330, Japan
        }%
    \affiliation{
    Research Center for Materials Nanoarchitectonics (MANA), 
                National Institute for Materials Science (NIMS), 
        Namiki 1-1, Tsukuba 305-0044, Japan
        }%
    \affiliation{
                Center for Spintronics Research Network (CSRN), 
        Osaka University, Toyonaka 560-8531, Japan 
        }%

\date{\today}

\begin{abstract}
We investigate the bulk photovoltaic effect in monolayer SnS using an
 effective tight-binding model derived from first-principles
 calculations. By comparing short-range and long-range hopping models,
 we show that the essential features of the shift current conductivity
 are captured by a minimal model. The shift current is decomposed into
 transition intensity and shift vector, enabling identification of
 dominant interband transitions. 
 The comparison reveals that long-range hopping processes quantitatively modify the peak positions and magnitudes, while the short-range model retains the characteristic low-energy structure of the nonlinear response.
 Our findings provide a transparent framework for understanding and designing bulk photovoltaic effects in two-dimensional materials. 
\end{abstract}


\maketitle

\section{Introduction}
The bulk photovoltaic effect (BPVE) has emerged as a promising mechanism for light-to-electricity conversion beyond the conventional p--n junction paradigm, as it enables the generation of photocurrent in non-centrosymmetric materials without requiring built-in electric fields \cite{Fridkin2001,Sipe2000,Young2012,Morimoto2016_SciAdv}. In particular, the shift current, one of the dominant contributions to BPVE, originates from the real-space displacement of electron wave packets during optical excitation and is closely related to the geometric properties of Bloch wavefunctions \cite{Sipe2000,Morimoto2016_SciAdv,Morimoto2016_PRB,Fregoso2017}. This intrinsic mechanism has attracted considerable attention for its potential to exceed the Shockley--Queisser limit and for its robustness against scattering processes \cite{Shockley1961,Young2012,Spanier2016,Cook2017}.

Recent advances in two-dimensional (2D) materials have opened new opportunities for realizing efficient BPVE. In contrast to bulk ferroelectrics, several 2D systems exhibit strong in-plane ferroelectric polarization, which is less susceptible to depolarization effects when interfaced with electrodes \cite{Wu2016,Ding2017,Fei2016}. Among them, group-IV monochalcogenides such as SnS have attracted particular interest due to their puckered crystal structures, strong in-plane ferroelectricity, and pronounced electronic anisotropy \cite{Qiao2014,Fei2016,Rangel2017}. Experimentally, monolayer SnS has been successfully synthesized and shown to exhibit purely in-plane ferroelectricity at room temperature, confirmed by piezoresponse force microscopy \cite{grows_SnS,kawamoto,bao2019}. These features make monolayer SnS a promising platform for investigating nonlinear optical responses and for designing next-generation optoelectronic devices.

While first-principles calculations have provided valuable insights into BPVE in various materials, they often offer limited physical transparency regarding the microscopic origin of the shift current \cite{Young2012,Tan2016,Rangel2017}. In particular, it remains challenging to identify which interband transitions and regions in momentum space dominantly contribute to the photocurrent, and how these contributions are governed by the underlying electronic structure. For the purpose of material design, it is therefore desirable to construct simplified yet quantitatively reliable models that can elucidate the essential mechanisms of BPVE in an intuitive manner \cite{Cook2017}.

In this work, we develop an effective tight-binding framework for monolayer SnS based on Wannier function interpolation of first-principles calculations, and use it to systematically investigate its nonlinear optical response. By constructing both short-range (SR) and long-range (LR) hopping models, we demonstrate that the essential features of the shift current conductivity can be captured within a minimal model, while extended hopping improves quantitative accuracy. Furthermore, we provide an intuitive decomposition of the shift current in terms of the transition intensity and the shift vector, enabling a clear identification of the dominant contributions in both energy and momentum space.

Our analysis clarifies how specific interband transitions near the band gap contribute to the observed spectral features and how these contributions are modified by the hopping range in the effective model. These findings establish a transparent connection between electronic structure, symmetry, and nonlinear optical response, and provide practical design principles for optimizing BPVE in low-dimensional materials.

This paper is organized as follows. 
In Sec.~II, we present the electronic structure of monolayer SnS and construct effective tight-binding models. 
In Sec.~III, we discuss the linear optical response as a benchmark for validating the model. 
In Sec.~IV, we formulate the shift current conductivity and introduce its decomposition in terms of physically intuitive quantities, followed by a detailed analysis of the calculated results. 
Finally, Sec.~V summarizes our findings.
In Appendix~\ref{app:hamiltonian}, we provide the explicit form of the Hamiltonian for the SR model.
In Appendix~\ref{app:shift_derivation}, we provide a detailed derivation of the shift current conductivity within the length-gauge and velocity-gauge formalisms.
In Appendix~\ref{app:shift_vector}, we present contour plots of the shift vector and transition intensity, which provide an intuitive understanding of the shift current conductivity in momentum space.

\begin{figure*}
  \begin{center}
    \includegraphics[width=\textwidth]{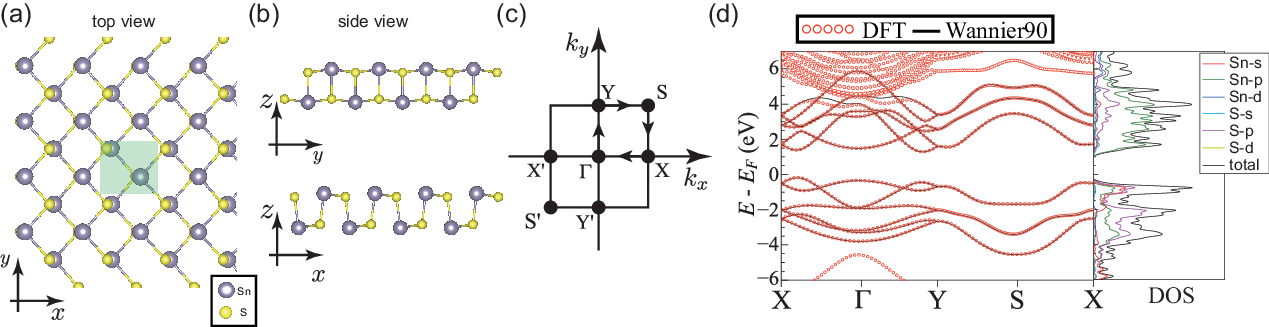}
    \caption{Crystal structure and Brillouin zone of monolayer SnS. 
    (a) Top view of the lattice, where the green rectangle denotes the
   primitive unit cell containing two Sn atoms and two S atoms.  
    (b) Side views along the armchair and zigzag directions,
   highlighting the puckered geometry that breaks inversion symmetry and
   gives rise to in-plane ferroelectric polarization.  
    (c) First Brillouin zone with high-symmetry points $\Gamma$, $X$,
   $Y$, and $S$ used in the band structure calculations.
(d) Energy band structure (left panel) and DOS (right panel) of monolayer SnS along 
    high-symmetry points in the first Brillouin zone.
    Red circles represent the DFT results, while black lines denote the Wannier-based tight-binding model.}
    \label{fig:SnS}
  \end{center}
\end{figure*}

\section{Electronic Structure of Monolayer SnS}
\label{sec:electronic_structure}
\subsection{Crystal structure}
Monolayer tin sulfide (SnS) is a two-dimensional semiconductor belonging
to the group-IV monochalcogenides (XS, XSe; X = Sn, Ge), which have
attracted considerable interest due to their intrinsic ferroelectricity
and strong in-plane anisotropy \cite{Fei2016,Rangel2017}.  
In contrast to conventional ferroelectrics, monolayer SnS exhibits
robust in-plane polarization originating from its puckered lattice
structure along the armchair direction, making it particularly suitable
for nanoscale device applications.

The crystal structure of monolayer SnS is shown in Fig.~\ref{fig:SnS}.  
The unit cell consists of two Sn atoms and two S atoms arranged in a
rectangular lattice. The puckered configuration breaks inversion
symmetry and leads to a finite spontaneous polarization within the
plane. 
Due to this puckered structure, monolayer SnS retains a large
spontaneous polarization even at elevated temperatures, making it a
promising candidate for ferroelectric and optoelectronic applications
\cite{Fei2016,SnS_Curie}.

\subsection{Effective tight-binding model}
\label{sec:tb_model}
The electronic structure of monolayer SnS was investigated within the
framework of density functional theory (DFT) using the Vienna \textit{Ab
initio} Simulation Package (VASP)~\cite{Kresse1996,Kresse1999}.  
The projector augmented-wave (PAW) method~\cite{Blochl1994} was
employed, and the exchange--correlation functional was treated within
the generalized gradient approximation (GGA) in the
Perdew--Burke--Ernzerhof (PBE) form~\cite{Perdew1996}.  
To avoid spurious interactions between periodic images, a vacuum region
larger than $15~\mathrm{\AA}$ was introduced along the out-of-plane
direction. 
The plane-wave cutoff energy was set to $440~\mathrm{eV}$, and the
Brillouin zone (BZ) was sampled using a $\Gamma$-centered
Monkhorst--Pack grid of $16 \times 16 \times 1$~\cite{Monkhorst1976}.  
The total energy convergence criterion was set to $10^{-10}$~eV.

To analyze the low-energy electronic structure relevant to optical
transitions, we construct an effective tight-binding model based on
maximally localized Wannier functions
(MLWFs)~\cite{Marzari1997,Souza2001,Marzari2012}.  
The Wannier functions were generated using the \textsc{Wannier90}
code~\cite{Mostofi2008}, starting from the DFT band structure.  
The resulting Wannier-interpolated Hamiltonian provides an accurate
description of the bands near the band gap while significantly reducing
the computational cost for evaluating optical responses. 

Figure~\ref{fig:SnS}(d) shows the density of states (DOS) and the
comparison between the DFT band structure and the Wannier-based
tight-binding model.  
The excellent agreement demonstrates that the constructed model
faithfully reproduces the essential features of the electronic structure
in the energy window of interest. 

The constructed tight-binding model accurately captures the band gap and
symmetry-protected degeneracies along the $Y \to S \to X$ path in the
Brillouin zone.  
The effective Hamiltonian is represented by a $12 \times 12$ matrix, reflecting the $p$-orbital manifold of Sn and S atoms that dominates the low-energy electronic structure. 
This Wannier-based formulation enables an efficient and physically
transparent analysis of optical and nonlinear response functions in
monolayer SnS. The details of the effective tight-binding models with different hopping ranges are described in Appendix~\ref{app:hamiltonian}.

\section{Linear Optical Response of Monolayer SnS}
\label{sec:linear_response}
\begin{figure}
  \begin{center}
    \includegraphics[width=0.45\textwidth]{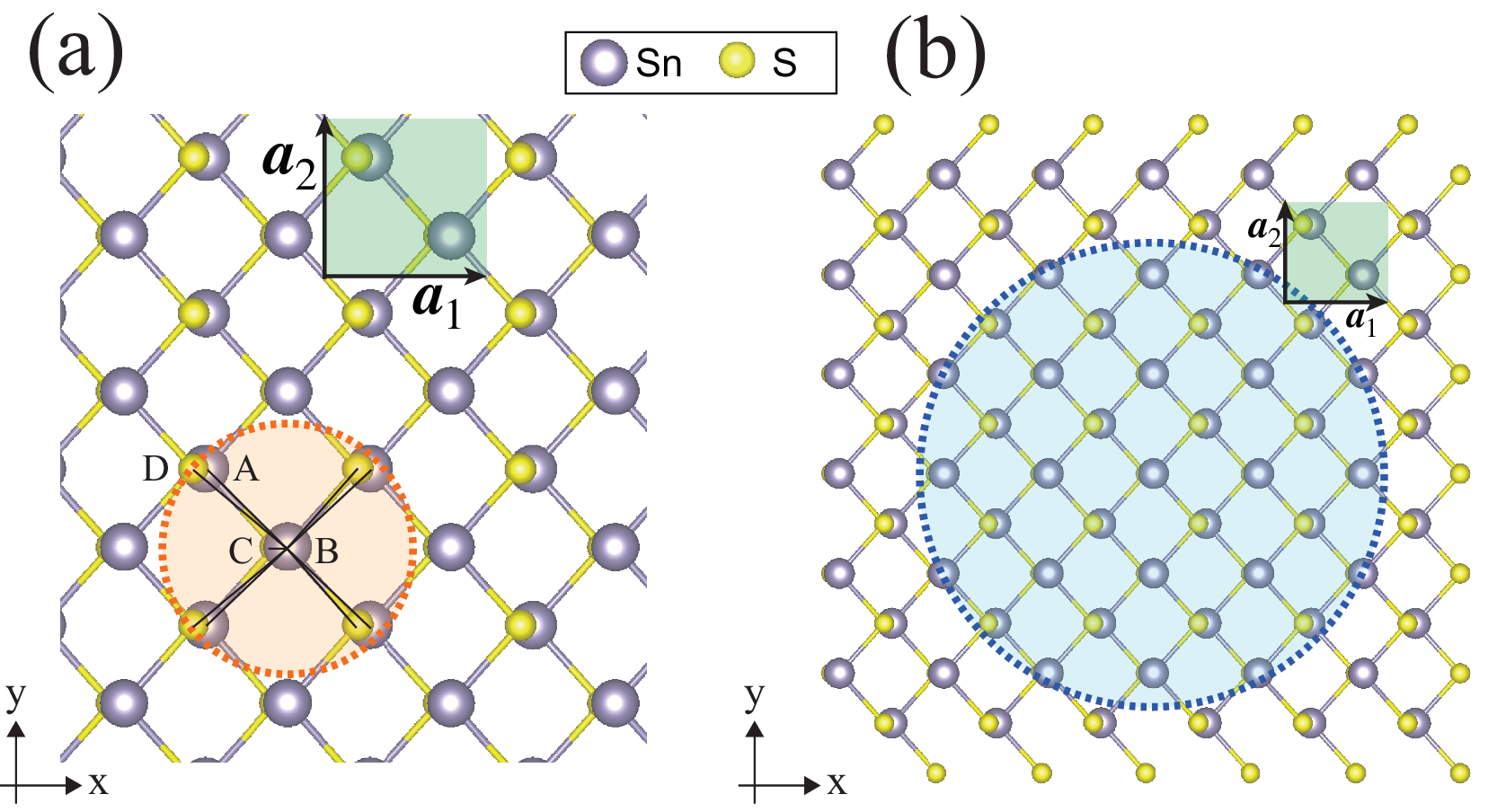}
    \caption{
    Tight-binding models with different hopping ranges. 
    (a) Short-range (SR) model including only nearest-neighbor hopping. 
    (b) Long-range (LR) model including extended hopping processes. 
    The green regions indicate the unit cells, and $\bm{a}_1$ and $\bm{a}_2$ denote the lattice vectors.}
    \label{fig:2model_structure}
  \end{center}
\end{figure}

In this section, we investigate the linear optical response of monolayer SnS using the effective tight-binding models introduced in Sec.~\ref{sec:tb_model}. 
To balance computational efficiency and accuracy, we consider two models with different hopping ranges: a short-range (SR) model including only nearest-neighbor hopping terms, and a long-range (LR) model incorporating extended hopping processes, as illustrated in Fig.~\ref{fig:2model_structure}.

\begin{figure}[h]
  \begin{center}
    \includegraphics[width=0.45\textwidth]{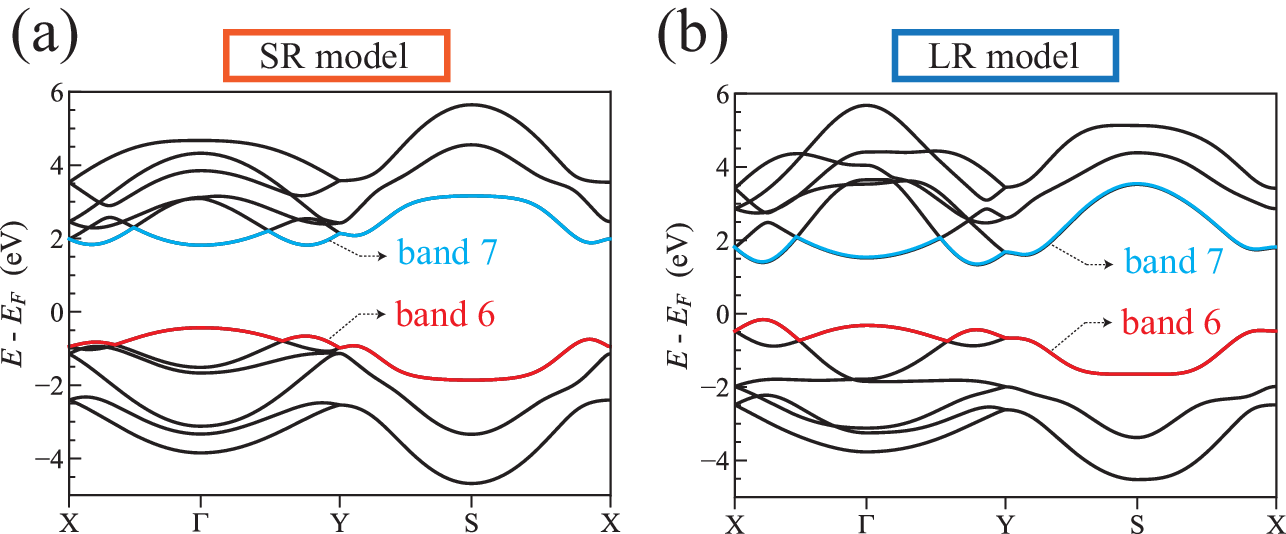}
    \caption{
    Band structures of monolayer SnS obtained from (a) the SR model and (b) the LR model. 
    The SR model captures the essential band gap and symmetry-protected degeneracies along the $Y \to S \to X$ path, while the LR model improves quantitative accuracy.}
    \label{fig:2model_bands}
  \end{center}
\end{figure}

Figure~\ref{fig:2model_bands} shows the corresponding band structures obtained from the two models. 
The LR model reproduces the dispersion near the band edges more accurately, while the SR model captures the essential low-energy features relevant to optical transitions.

When an external electromagnetic field is applied, the induced current density is described within linear response theory as
\begin{equation}
  J_i^{(1)}(\omega) = \sum_j \sigma_{ij}(\omega) E_j(\omega),
\end{equation}
where $E_j(\omega)$ is the electric field component. 
The linear optical conductivity $\sigma_{ij}(\omega)$ is calculated using the Kubo formula~\cite{kubo,Sipe2000}:
\begin{widetext}
\begin{align}
  \sigma_{ij}(\omega) = \frac{i\hbar e^2}{S} \sum_{\bm{k}} \sum_{n,m}
  \frac{f(E_{n\bm{k}}) - f(E_{m\bm{k}})}{E_{m\bm{k}} - E_{n\bm{k}}}
 \cdot \frac{\langle u_{n\bm{k}} | \hat{v}_i | u_{m\bm{k}} \rangle 
        \langle u_{m\bm{k}} | \hat{v}_j | u_{n\bm{k}} \rangle}
       {E_{m\bm{k}} - E_{n\bm{k}} - \hbar\omega - i\eta},
  \label{eq:kubo}
\end{align}
\end{widetext}
where $|u_{n\bm{k}}\rangle$ is the cell-periodic Bloch function, and $f(E_{n\bm{k}})$ is the Fermi--Dirac distribution.

The velocity operator is given by
\begin{equation}
  \hat{\bm{v}}_{\bm{k}} = \frac{1}{\hbar} \nabla_{\bm{k}} \hat{H}(\bm{k}).
\end{equation}

In the static limit, the transverse conductivity is related to the Berry curvature~\cite{Xiao2010}:
\begin{equation}
  \sigma_{xy} = \frac{e^2}{\hbar} \frac{1}{S} \sum_{\bm{k}} \Omega^z(\bm{k}),
\end{equation}
where the Berry curvature is defined as
\begin{widetext}
\begin{align}
  \Omega^z(\bm{k}) = -2\hbar^2\,\mathrm{Im} \sum_n f(E_{n\bm{k}})
\sum_{m \ne n}
\frac{\langle u_{n\bm{k}} | \hat{v}_x | u_{m\bm{k}} \rangle
        \langle u_{m\bm{k}} | \hat{v}_y | u_{n\bm{k}} \rangle}
       {(E_{m\bm{k}} - E_{n\bm{k}})^2}.
\end{align}
\end{widetext}

\begin{figure*}
  \begin{center}
    \includegraphics[width=0.9\textwidth]{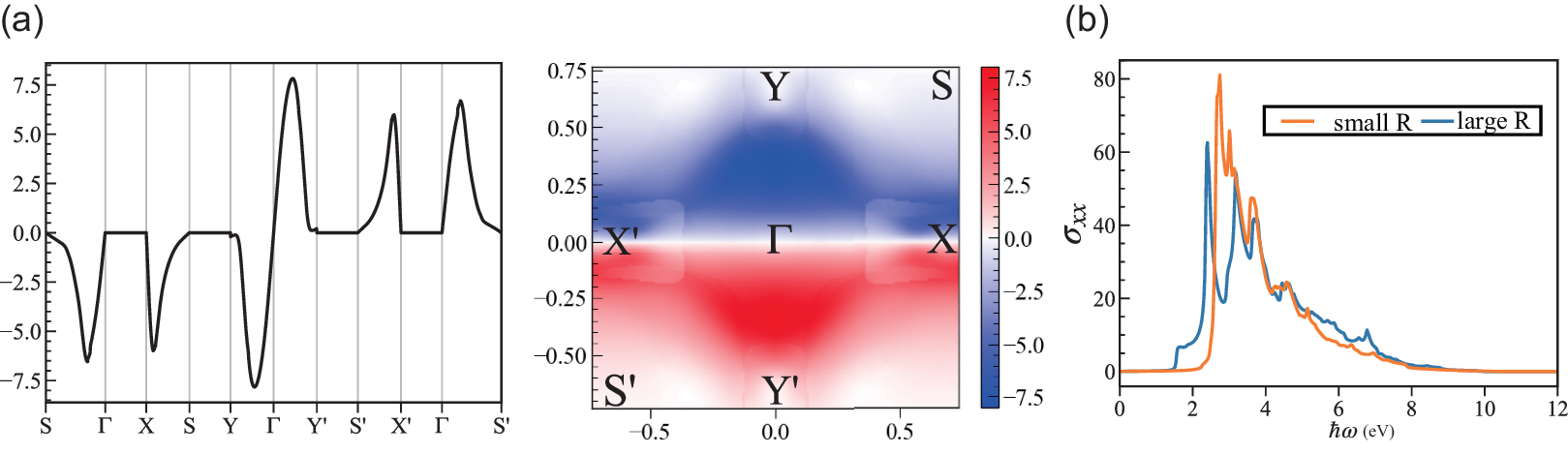}
    \caption{
    Berry curvature and optical conductivity of monolayer SnS. 
    (a) Berry curvature $\Omega^z(\bm{k})$ in the first Brillouin zone calculated using the LR model. 
    The anisotropic distribution reflects inversion symmetry breaking. 
    (b) Longitudinal optical conductivity $\sigma_{xx}(\omega)$ for the SR and LR models. 
    The LR model provides quantitative accuracy, while the SR model captures essential features.}
    \label{fig:berry}
  \end{center}
\end{figure*}
Although the Berry curvature is locally finite due to inversion symmetry breaking, its integral over the Brillouin zone vanishes, resulting in $\sigma_{xy} = 0$. 
This is consistent with time-reversal symmetry, which forbids an intrinsic Hall response in monolayer SnS~\cite{Xiao2010}.

Figure~\ref{fig:berry} summarizes the Berry curvature and the linear optical response of monolayer SnS. 
Figure~\ref{fig:berry}(a) shows the Berry curvature \(\Omega^z(\bm{k})\) calculated using the LR model. 
The Berry curvature exhibits an anisotropic distribution in the first Brillouin zone, reflecting the reduced symmetry of monolayer SnS. 
Figure~\ref{fig:berry}(b) compares the longitudinal optical conductivity \(\sigma_{xx}(\omega)\) obtained from the SR and LR models. 
The SR model reproduces the main low-energy spectral features of the LR model, while the inclusion of longer-range hoppings modifies the peak positions and detailed spectral shapes. 
In particular, the peak heights in the linear optical conductivity are relatively similar between the two models compared with the shift-current spectra discussed in Sec.~IV.B.

The linear response formulation provides the basis for analyzing the nonlinear optical response, including the bulk photovoltaic effect, which is discussed in the following section~\cite{Morimoto2016_PRB,Rangel2017}.

\section{Shift Current in Monolayer SnS}
\label{sec:shift_current}
\subsection{Formalism}
In this section, we investigate the bulk photovoltaic effect in monolayer SnS, focusing on the shift current contribution. 
The shift current originates from the real-space displacement of electron wave packets during optical excitation and is governed by the geometric properties of Bloch states.

We employ the standard length-gauge formulation of nonlinear optical response developed in Refs.~\cite{Aversa1995,Sipe2000,Morimoto2016_PRB}. 
The detailed derivation of the shift current conductivity is provided in Appendix~\ref{app:shift_derivation}. 
Here, we present the final expression and its physical interpretation.

The shift current conductivity is expressed as
\begin{widetext}
\begin{equation}
  \sigma_{ijk}^{\mathrm{shift}}(0;\omega,-\omega)
  = -\frac{i\hbar e^3}{2S} \sum_{\bm{k}} \sum_{n,m}
  \frac{f_{nm}}{E_{mn}^2(E_{mn} - \hbar\omega - i\eta)}
  \left(\alpha_{mn}^{jk}(\bm{k})\,R_{nm}^{ik}(\bm{k})
       +\alpha_{mn}^{kj}(\bm{k})\,R_{nm}^{ij}(\bm{k})\right),
  \label{eq:shift_current}
\end{equation}
\end{widetext}
where $E_{mn} = E_{m\bm{k}} - E_{n\bm{k}}$, $f_{nm} = f(E_{n\bm{k}}) - f(E_{m\bm{k}})$,
and $S$ is the unit cell area.
The transition intensity is defined as
\begin{equation}
  \alpha_{mn}^{jk}(\bm{k}) = v_{mn}^j v_{nm}^k,
\end{equation}
and $R_{nm}^{ij}(\bm{k})$ is the shift vector.

The shift vector is given by
\begin{equation}
  R_{nm}^{ij}(\bm{k}) = \partial_{k_i} \phi_{nm}^j + \xi_{nn}^i - \xi_{mm}^i,
\end{equation}
where $\phi_{nm}^j$ is the phase of the velocity matrix element $v_{nm}^j$,
and $\xi_{mm}^i = i\langle u_{m\bm{k}}|\partial_{k_i}|u_{m\bm{k}}\rangle$ is the Berry connection.

Equation~(\ref{eq:shift_current}) provides a transparent physical picture of the shift current.
The nonlinear response is determined by the product of the transition intensity $\alpha_{mn}^{jk}(\bm{k})$ and the shift vector $R_{nm}^{ij}(\bm{k})$,
which represents the real-space displacement of the electron wave packet during optical excitation.

In monolayer SnS, the absence of inversion symmetry allows for a finite shift current response. 
In addition, the crystal symmetry restricts the nonzero components of the shift-current conductivity tensor. 
For linearly polarized light incident from the \(z\)-direction, the symmetry-allowed components are 
\(\sigma^{\rm shift}_{xxx}\), \(\sigma^{\rm shift}_{xyy}\), and 
\(\sigma^{\rm shift}_{yxy}=\sigma^{\rm shift}_{yyx}\), whereas the other components vanish. 
In the following, we analyze these symmetry-allowed components based on the effective tight-binding models.

\begin{figure}[h]
  \begin{center}
    \includegraphics[width=0.45\textwidth]{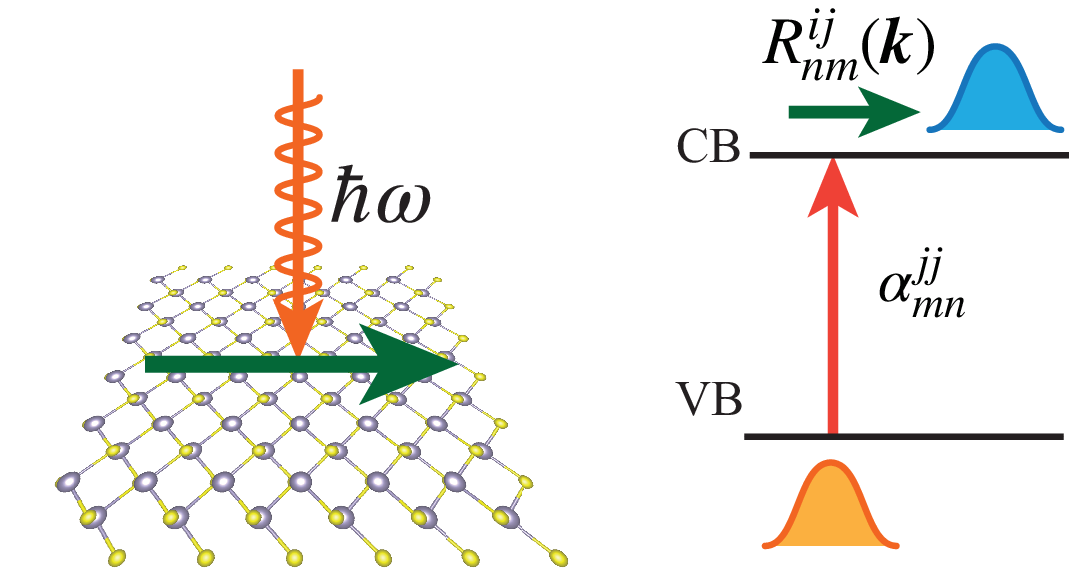}
\caption{
Schematic illustration of the shift current mechanism in monolayer SnS. 
(a) Real-space picture of photoexcited carrier displacement. 
(b) Interband transition between valence and conduction bands, 
characterized by the transition intensity and shift vector.}
    \label{fig:shift_current}
  \end{center}
\end{figure}
Figure~\ref{fig:shift_current} schematically illustrates the shift-current mechanism in monolayer SnS in terms of the transition intensity and shift vector.
The transition intensity describes the strength of the optical transition, while the shift vector represents the real-space displacement of the electron wave packet during optical excitation.
In the following subsection, we evaluate the symmetry-allowed components using the SR and LR tight-binding models and analyze how the hopping range modifies the shift-current spectra.

The comparison between the SR and LR models reveals that the essential features of the shift current are already captured by the SR model, 
while the LR model provides quantitative corrections. 
This indicates that the shift current in monolayer SnS is primarily governed by the low-energy electronic structure.

These results establish a clear connection between electronic structure, symmetry, and nonlinear optical response, 
and provide useful insights for designing materials with enhanced bulk photovoltaic effects.

\subsection{Shift Current Conductivity in Monolayer SnS}
We calculate the shift current conductivity of monolayer SnS under linearly polarized light incidence. 
Monolayer SnS belongs to the $Pmn2_1$ space group.
Due to its symmetry, the shift current conductivity components $\sigma^{\rm{shift}}_{xxx}$, $\sigma^{\rm{shift}}_{xyy}$, 
and $\sigma^{\rm{shift}}_{yxy} = \sigma^{\rm{shift}}_{yyx}$ are finite when linearly polarized light is incident from the $z$-direction. 
All other components vanish. 
\begin{figure*}
  \begin{center}
    \includegraphics[width=170mm]{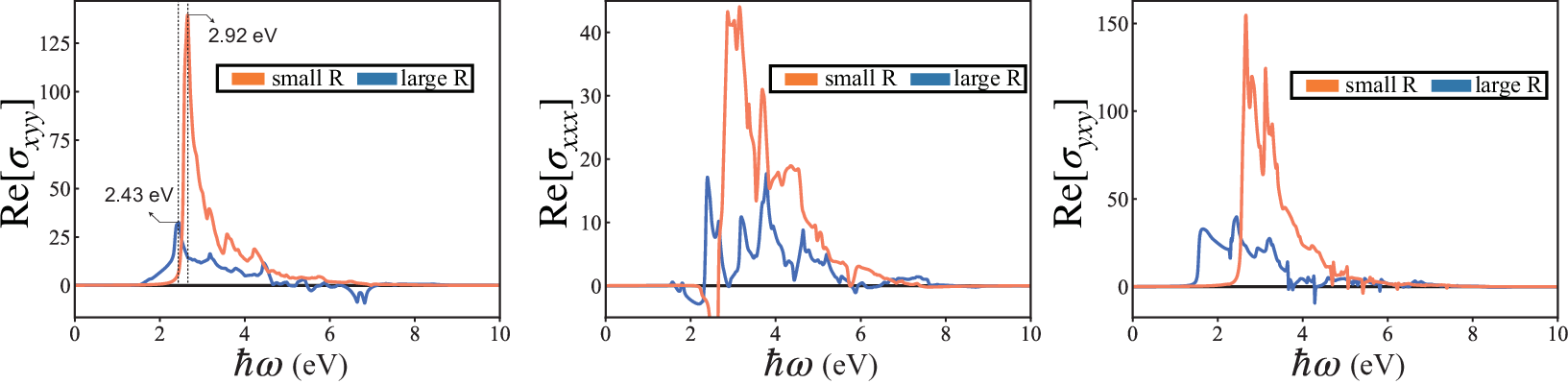}
    \caption{The real part of shift current conductivities $\sigma^{\rm{shift}}_{xyy}$ (left panel), 
    $\sigma^{\rm{shift}}_{xxx}$ (center panel) 
    and $\sigma^{\rm{shift}}_{yxy}$ (right panel) for monolayer SnS. 
    The unit of shift current conductivity is $e^3/\hbar$.}
    \label{fig:shift3paturn}
  \end{center}
\end{figure*}
Figure~\ref{fig:shift3paturn} shows the calculated shift current conductivity obtained using both the SR and LR models. 
It is evident that all three shift current conductivity components exhibit higher peak values in the SR model.

To clarify the origin of the different peak magnitudes between the SR and LR models, we further analyze the momentum-resolved contribution to the shift-current conductivity \(\sigma^{\rm shift}_{xyy}\) at the representative low-energy peaks in the left panel of Fig.~\ref{fig:shift3paturn}.
\begin{figure}
  \begin{center}
    \includegraphics[width=85mm]{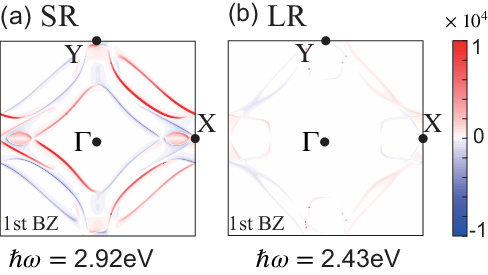}
    \caption{Momentum-resolved contribution to \({\rm Re}[\sigma^{\rm shift}_{xyy}]\) at the representative low-energy peaks of Fig.~\ref{fig:shift3paturn}. 
(a) and (b) show the SR and LR models evaluated at \(\hbar\omega=2.92~{\rm eV}\) and \(2.43~{\rm eV}\), respectively. 
The plotted quantity is the shift-current integrand before Brillouin-zone summation, including the same resonance denominator and broadening factor as in the spectral calculation, with the overall prefactor omitted.}
    \label{fig:integrand_map_xyy}
  \end{center}
\end{figure}
Figure~\ref{fig:integrand_map_xyy} shows the resulting \(k\)-resolved contribution to 
\({\rm Re}[\sigma^{\rm shift}_{xyy}]\) before the Brillouin-zone summation, evaluated at 
\(\hbar\omega=2.92~{\rm eV}\) for the SR model and \(\hbar\omega=2.43~{\rm eV}\) for the LR model. 
The SR model exhibits pronounced local contributions along the resonant regions of the Brillouin zone. 
In contrast, the corresponding contributions in the LR model are less pronounced and more broadly distributed. 
The stronger local contributions in the SR model are consistent with the larger peak observed in the shift-current spectrum.
This indicates that the SR--LR difference is governed by the momentum-space distribution of the full shift-current integrand, 
including the transition matrix elements, the shift-vector-related numerator, and the resonance denominator, rather than by the band-energy structure alone.
This behavior contrasts with the linear optical response, where the shift-vector-related numerator is absent, and highlights the sensitivity of the shift-current response to the detailed momentum-space structure of the electronic states.
Therefore, the SR model captures the qualitative low-energy features of the shift-current response, whereas the LR model is necessary for a more quantitatively reliable description.

For completeness, Appendix~\ref{app:shift_vector} presents band-resolved contour plots of the shift vector \(R^{ij}_{nm}(\bm{k})\) and transition intensity \(\alpha^{jk}_{mn}=v^j_{mn}v^k_{nm}\) for selected interband transitions near the band gap. 
The band indices are defined by the ascending order of the eigenvalues in the band structure shown in Fig.~\ref{fig:2model_bands}.

\section{Conclusion}\label{Conclusion}
In this study, we theoretically investigated the electronic structure
and optical properties of monolayer SnS. Two effective tight-binding
models with different hopping ranges were constructed using maximally
localized Wannier functions to reproduce the energy band structure near
the band gap. For the first-order optical response, we confirmed that
monolayer SnS exhibits finite longitudinal conductivity, 
validating the SR model. 
We also demonstrated that Berry curvature originates from
inversion symmetry breaking and reverses sign under wavevector
inversion, resulting in vanishing transverse conductivity. For the
second-order nonlinear optical response, we showed that monolayer SnS
generates a shift current under linearly polarized light without
external strain, unlike centrosymmetric materials that require strain
for BPVE. 
Furthermore, our simplified SR model successfully captures the
essential features of shift current conductivity. 
These findings provide a transparent framework for understanding shift
current responses in monolayer SnS and related non-centrosymmetric
2D materials.

\section*{ACKNOWLEDGMENTS}
This work was supported by JSPS KAKENHI (Grants No. JP25K01609,
No. JP22H05473, and No. JP21H01019), JST CREST (Grant
No. JPMJCR19T1). K. W. acknowledges the financial support for Basic
Science Research Projects (Grant No. 2401203) from the Sumitomo
Foundation, and the special individual research support from Kwansei Gakuin University.
T. K. acknowledges the Sasakawa Scientific Research Grant from The Japan Science Society.


\appendix
\section{Details of the Hamiltonian in the SR model}\label{app:hamiltonian}
\begin{figure*}
  \begin{center}
    \includegraphics[width=0.6\textwidth]{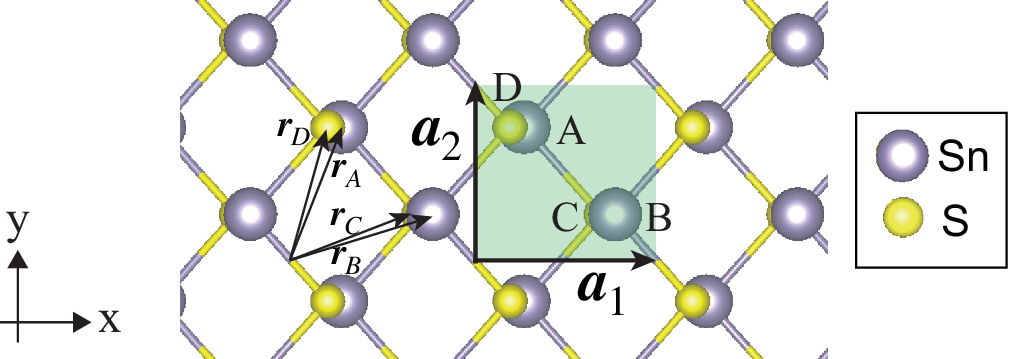}
    \caption{
    Top view of the lattice structure of monolayer SnS. 
    The shaded rhombus indicates the primitive unit cell containing four nonequivalent sites (A–D).}
    \label{lattice_SnS}
  \end{center}
\end{figure*}

Here, we present the Hamiltonian of the short-range (SR) tight-binding model for monolayer SnS. 
Figure~\ref{lattice_SnS} shows the lattice structure. 
The shaded rhombus denotes the primitive unit cell, which contains four nonequivalent atomic sites labeled A, B, C, and D.

The primitive lattice vectors are given by $\bm{a}_1 = (4.28, 0)$ and $\bm{a}_2 = (0, 4.08)$. 
The position vectors of the atomic sites are defined as 
$\bm{r}_A = (0.473, 3.06)$, $\bm{r}_B = (2.61, 1.02)$, 
$\bm{r}_C = (2.29, 1.02)$, and $\bm{r}_D = (0.15, 3.06)$, respectively.

In monolayer SnS, the electronic states near the band gap are mainly derived from the $p$ orbitals of Sn and S atoms. 
Since the unit cell contains two Sn atoms and two S atoms, the effective tight-binding model is constructed from 12 Wannier orbitals, 
corresponding to three orbitals per atomic site, resulting in a $12 \times 12$ Hamiltonian.

The SR model includes only nearest-neighbor hopping terms. 
The eigenvalue equation is given by
\begin{equation}
  H(\bm{k})\ket{\psi_{n\bm{k}}} = E_{n\bm{k}}\ket{\psi_{n\bm{k}}},
\end{equation}
where $H(\bm{k})$ is the Bloch Hamiltonian at wavevector $\bm{k} = (k_x, k_y)$, 
and $E_{n\bm{k}}$ is the eigenenergy with band index $n = 1,2,\ldots,12$.

The matrix elements of $H(\bm{k})$ are written as
\begin{widetext}
\begin{equation}
  h_{\alpha \beta}(\bm{k}) =
  \begin{cases}
    \gamma_{\alpha \beta} \, e^{i\bm{k}\cdot\bm{\tau}_{\alpha \beta}}, 
    & (\alpha, \beta) \in \{(A,D),(D,A),(B,C),(C,B)\}, \\
    \gamma_{\alpha \beta} \left(e^{i\bm{k}\cdot\bm{\tau}_{\alpha \beta}} + e^{-i\bm{k}\cdot\bm{\tau}_{\alpha \beta}}\right), 
    & \text{otherwise},
  \end{cases}
\end{equation}
\end{widetext}
where $\gamma_{\alpha\beta}$ (in eV) is the hopping parameter between orbitals $\alpha$ and $\beta$, as listed in Table~\ref{hopping_energy}. 
The relative displacement vector between sites is defined as 
$\bm{\tau}_{\alpha\beta} = \bm{r}_\beta - \bm{r}_\alpha$.

This formulation provides a minimal yet physically transparent model that captures the essential low-energy electronic structure relevant to optical transitions.

\begin{table*}
  \caption{
Hopping parameters $\gamma_{\alpha\beta}$ of the short-range (SR) tight-binding model for monolayer SnS (in eV).
The indices $A$, $B$, $C$, and $D$ denote atomic sites, and the subscripts $1$, $2$, and $3$ correspond to the three $p$ orbitals at each site.}
  \label{hopping_energy}
\begin{ruledtabular}
\begin{tabular}{l*{12}{c}}
 & ${A_1}$ & ${A_2}$ & ${A_3}$ & ${B_1}$ & ${B_2}$ & ${B_3}$ & ${C_1}$ & ${C_2}$ & ${C_3}$ & ${D_1}$ & ${D_2}$ & ${D_3}$ \\
\hline
${A_1}$ & -1.927 & -0.029 & -0.001 & -0.118 & 0.109 & -0.032 & -1.447 & 0.641 & 0.002 & 2.039 & 0.092 & 0.000 \\
${A_2}$ & -0.029 & -0.872 & -0.001 & -0.078 & 0.312 & -0.057 & 0.721 & 2.433 & -0.087 & -0.320 & -0.380 & 0.000 \\
${A_3}$ & -0.001 & -0.001 & 0.010 & -0.027 & 0.027 & -0.187 & -0.024 & -0.086 & 2.892 & 0.000 & 0.000 & -0.573 \\
${B_1}$ & -0.118 & -0.078 & -0.027 & -1.927 & 0.029 & -0.001 & 2.039 & -0.092 & 0.000 & -1.457 & -0.675 & 0.012 \\
${B_2}$ & 0.109 & 0.312 & 0.027 & 0.029 & -0.872 & 0.001 & 0.320 & -0.380 & 0.000 & -0.752 & 2.550 & 0.074 \\
${B_3}$ & -0.032 & -0.057 & -0.187 & -0.001 & 0.001 & 0.010 & 0.000 & 0.000 & -0.573 & -0.007 & 0.068 & 2.838 \\
${C_1}$ & -1.447 & 0.721 & -0.024 & 2.039 & 0.320 & 0.000 & -3.918 & 0.038 & 0.000 & 0.565 & -0.191 & -0.028 \\
${C_2}$ & 0.641 & 2.433 & -0.086 & -0.092 & -0.380 & 0.000 & 0.038 & -4.333 & -0.001 & 0.241 & 0.407 & -0.016 \\
${C_3}$ & 0.002 & -0.087 & 2.892 & 0.000 & 0.000 & -0.573 & 0.000 & -0.001 & -4.170 & -0.037 & -0.018 & 0.243 \\
${D_1}$ & 2.039 & -0.320 & 0.000 & -1.457 & -0.752 & -0.007 & 0.565 & 0.241 & -0.037 & -3.918 & -0.038 & 0.000 \\
${D_2}$ & 0.092 & -0.380 & 0.000 & -0.675 & 2.550 & 0.068 & -0.191 & 0.407 & -0.018 & -0.038 & -4.333 & 0.001 \\
${D_3}$ & 0.000 & 0.000 & -0.573 & 0.012 & 0.074 & 2.838 & -0.028 & -0.016 & 0.243 & 0.000 & 0.001 & -4.170 \\
\end{tabular}
\end{ruledtabular}
\end{table*}

\section{Derivation of Shift Current Conductivity}
\label{app:shift_derivation}
In this appendix, we derive the shift current conductivity 
$\sigma_{ijk}^{\mathrm{shift}}(0;\omega,-\omega)$ 
within the length-gauge formalism and the velocity-gauge formalism, following 
Refs.~\cite{Aversa1995,Sipe2000,Morimoto2016_PRB}.

\subsection{Second-order optical response}
We consider the interaction of a crystalline solid with a monochromatic light field in the length gauge. 
The perturbation Hamiltonian is given by
\begin{equation}
  H'(t) = e \sum_j E_j(t)\,\hat{r}_j,
\end{equation}
where $\hat{r}_j$ is $j$ element of the position operator $\hat{\bm{r}}$ and 
$E_j(t)=E_j e^{-i\omega t}+E_j^* e^{i\omega t}$.

The dynamics of the density matrix 
$\hat{\rho}=\hat{\rho}^{(0)}+\hat{\rho}^{(1)}+\hat{\rho}^{(2)}+\cdots$
are governed by the von~Neumann equation
\begin{equation}
  i\hbar\frac{\partial\hat{\rho}}{\partial t}
  = \bigl[H_0+H'(t),\,\hat{\rho}\bigr].
\end{equation}

At equilibrium, $\hat{\rho}^{(0)}$ is given by the Fermi--Dirac distribution. 
Solving perturbatively, the first-order density matrix element in the band basis is
\begin{equation}
  \rho^{(1)}_{nm}(\omega)
  = \frac{e\,f_{nm}\,r^j_{nm}\,E_j(\omega)}
         {E_{mn}-\hbar\omega-i\eta},
\end{equation}
where $f_{nm}=f(E_{n\bm{k}})-f(E_{m\bm{k}})$ and \mbox{ $r^j_{nm}=\bra{u_{n\bm{k}}}\hat{\bm{r}}\cdot\bm{e}_j\ket{u_{m\bm{k}}}$ } is the matrix element of the $\hat{r}_j$ between states $n$ and $m$.
Similarly, the second-order density matrix is obtained as
\begin{widetext}
\begin{align}
\rho^{(2)}_{nm}(\omega+\omega';\omega,\omega')
=
\frac{e^2}{E_{mn}-\hbar(\omega+\omega')+i\eta}
\sum_l
\left(
\frac{f_{lm} r^{k}_{nl} r^{j}_{lm}}
{E_{ml}-\hbar\omega+i\eta}
-
\frac{f_{nl} r^{k}_{nl} r^{j}_{lm}}
{E_{ln}-\hbar\omega+i\eta}  
\right)E_j(\omega)E_k(\omega'),
\end{align}
\end{widetext}
where $\omega$ and $\omega^{\prime}$ are the frequencies of the incident light from $j$ and $k$ directions, respectively.

\subsection{Shift current conductivity}
For charge current operator $J$, the thermal expectation value of $J$ is given as
\begin{equation}
\langle J \rangle = \mathrm{Tr}(J \rho(t)).
\end{equation}
The DC second-order current is expressed as
\begin{equation}
  J_i^{(2)}
  = \sum_{jk}\sigma_{ijk}^{(2)}(0;\omega,\omega^{\prime})\,E_j(\omega)\,E_k(\omega^{\prime}),
\end{equation}
where $\sigma_{ijk}^{(2)}(0;\omega,\omega^{\prime})$ is the second-order conductivity tensor. In this paper, we focus on the photocurrent generated by BPVE, which corresponds to the case of $\omega' = -\omega$.
We can rewrite the second-order conductivity as
\begin{equation}
  \sigma_{ijk}^{(2)}(0;\omega,-\omega) = \sigma_{ijk}^{\rm{shift}}(0;\omega, -\omega) + \sigma_{ijk}^{\rm{injection}}(0;\omega, -\omega),
\end{equation}
where $\sigma_{ijk}^{\rm{shift}}$ and $\sigma_{ijk}^{\rm{injection}}$ are the shift and injection current contributions, respectively.
\begin{widetext}
Under linearly polarized light irradiation, the optical conductivities of the shift current 
$\sigma_{ijk}^{\rm{shift}}(0;\omega, -\omega)$ and the injection current $\sigma _{ijk} ^{\rm{injection}}(0;\omega, -\omega)$ are given by \cite{habara2023,NLO_spin,Sipe2000}
\begin{align}
  \sigma_{ijk}^{\mathrm{shift}}
  = \frac{e^3}{2\hbar S}\sum_{\bm{k}}\sum_{n,m}
    \frac{f_{nm}}{E_{mn}-\hbar\omega-i\eta}
    \left(r^j_{mn}r^k_{nm;i}+r^k_{mn}r^j_{nm;i}\right),
  \label{shift_length}
\end{align}
\begin{align}
  \sigma _{ijk} ^{\rm{injection}}(0;\omega, -\omega)  = \tau\frac {ie^3}{2\hbar S}\sum_{\bm{k}}
  \sum_{nm} \frac{f_{nm}}{E_{mn} - \hbar \omega - i \eta}\Delta^i_{nm}\{r^k_{mn},r^j_{nm}\},
  \label{inject_length}
\end{align}
\end{widetext}
where the generalized derivative $r^j_{nm;i}$ is expressed as
\begin{equation}
  \begin{aligned}
    r^{j}_{nm;i}
    =\frac{\partial{r^j_{nm}}}{\partial{k_i}}-i[\xi^i_{nn}(\boldsymbol{k})-\xi^i_{mm}(\boldsymbol{k})]r^j_{nm},
  \end{aligned}
  \label{gen_der2}
\end{equation}
where $\xi^i_{mm}$ is the Berry connection for the $m$-th band, given by
\begin{equation}
  \xi^i_{mm} = i\bra{u_{m\bm{k}}}\frac{\partial}{\partial k_i}\ket{u_{m\bm{k}}}.
\end{equation}

%

\subsection{Velocity representation and shift vector}
To evaluate the conductivity numerically, we express the position matrix elements in terms of velocity operators:
\begin{equation}
  r^j_{nm} = 
  \left\{
    \begin{aligned} 
      &\frac{v^j_{nm}}{i \omega_{nm}} \quad&(n \not= m) \\
      &0 \quad&(n=m),
    \end{aligned} 
  \right.
  \label{r^j_nm}
\end{equation}
where $v^j_{nm} = \bra{u_{n \bm {k}}} \hat{\bm{v}}\cdot\bm{e_j} \ket{u_{m \bm{k}}}$, $\Delta^j_{mn} = v^j_{mm} - v^j_{nn}$ and $\omega_{nm} = (E_{n\bm{k}} - E_{m\bm{k}})/\hbar$.
Here, we introduce the shift vector
\begin{equation}
  R^{ij}_{nm}(\bm{k}) = \partial_{k_i} \phi^j_{nm} + \xi^i_{nn} - \xi^i_{mm},
  \label{shift_vector_l}
\end{equation}
which includes the Berry connection difference. 
The phase term $\phi^j_{nm}$ corresponds to the phase of the group velocity matrix element as
\begin{equation}
  v^j_{nm} =
   \bra{u_{n\bm{k}}}\hat{\bm{v}}\cdot\bm{e}_j\ket{u_{m\bm{k}}}=|v^j_{nm}|{\rm
   e}^{\phi^j_{nm}}.
\end{equation}
From Eqs.~(\ref{gen_der2}) and (\ref{shift_vector_l}), we obtain
\begin{equation}
  r^j_{nm;i} = -i R^{ij}_{nm}(\bm{k}) r^j_{nm}.
\end{equation}
Substituting these expressions into the conductivity formula, Eq.~(\ref{shift_length}), we obtain the shift current conductivity in terms of the velocity matrix elements and the shift vector:
\begin{widetext}
\begin{equation}
  \sigma_{ijk}^{\mathrm{shift}}
  = -\frac{i\hbar e^3}{2S}\sum_{\bm{k}}\sum_{n,m}
    \frac{f_{nm}}{E_{mn}^2(E_{mn}-\hbar\omega-i\eta)}
    \left(v^j_{mn}v^k_{nm}R^{ik}_{nm}+v^k_{mn}v^j_{nm}R^{ij}_{nm}\right).
\end{equation}
\end{widetext}

Here, we consider velocity representation of the shift vector $R^{ij}_{nm}(\bm{k})$.
The generalized derivative of the position operator $r^{j}_{nm}$
can be expressed using the sum rule of differentiation, allowing the separation of the BPVE into 
shift and injection currents \cite{habara2023,NLO_spin,Sipe2000}.
The generalized derivative $r^{j}_{nm:i}$ is given by
\begin{widetext}
\begin{align}
  r^j_{nm;i} = \frac{r^i_{nm} \Delta^j_{mn} + r^j_{nm} \Delta^i_{mn}} {\omega_{nm}}
  + \frac{i}{\omega_{nm}} \sum_l (\omega_{lm}r^i_{nl}r^j_{lm} - \omega_{nl}r^j_{nl}r^i_{lm}),
  \label{gen_der1}
\end{align}


Using Eqs.~(\ref{gen_der1}) and (\ref{gen_der2}), we obtain 
\begin{equation}
  \begin{aligned}
    -iR^{ij}_{nm}(\bm{k})r^j_{nm}
    = \frac{r^i_{nm} \Delta^j_{mn} + r^j_{nm} \Delta^i_{mn}} {\omega_{nm}}
    + \frac{i}{\omega_{nm}} \sum_m (\omega_{lm}r^i_{nl}r^j_{lm} - \omega_{nl}r^j_{nl}r^i_{lm}).
    \label{iRr}
  \end{aligned}
\end{equation}
From Eq.~(\ref{r^j_nm}), the right-hand side of Eq.~(\ref{iRr}) can be rewritten as
\begin{equation}
  \begin{aligned}
    \frac{v^i_{nm}(v^j_{mm}-v^j_{nn})+ v^j_{nm}(v^i_{mm}-v^i_{nn})}{i\omega^2_{nm}}
    +\frac{i}{\omega_{nm}} \sum_l \Bigl (\omega_{lm} \frac{v^i_{nl}}{i\omega_{nl}} 
    \frac{v^j_{lm}}{i\omega_{lm}} - \omega_{nl} \frac{v^j_{nl}}{i\omega_{nl}} 
    \frac{v^i_{lm}}{i\omega_{lm}}\Bigr ) \\
    =\frac{v^i_{nm}(v^j_{mm}-v^j_{nn})+ v^j_{nm}(v^i_{mm}-v^i_{nn})}{i\omega^2_{nm}}
    + \frac{i}{\omega_{nm}} \sum_l\Bigl ( 
    -\frac{v^i_{nl} v^j_{lm}}{\omega_{nl}}   
    +\frac{v^j_{nl} v^i_{lm}}{\omega_{lm}} \Bigr ).
  \end{aligned}
\end{equation}
Thus, we obtain
\begin{equation}
  \begin{aligned}
    R^{ij}_{nm}(\bm{k}) = 
    -\frac{1}{v^j_{nm}}\Biggl [\frac{v^i_{nm}(v^j_{mm}-v^j_{nn})+ v^j_{nm}(v^i_{mm}-v^i_{nn})}{i\omega_{nm}}
    +i \sum_l\Biggl (-\frac{v^i_{nl} v^j_{lm}}{\omega_{nl}} +\frac{v^j_{nl} v^i_{lm}}{\omega_{lm}}\Biggr )\Biggr ]\\
    = \frac{i\hbar}{v^j_{nm}}\Biggl [\frac{v^i_{nm}(v^j_{mm}-v^j_{nn})+ v^j_{nm}(v^i_{mm}-v^i_{nn})}{E_{nm}}
    + \sum_l\Biggl (\frac{v^i_{nl} v^j_{lm}}{E_{nl}} -\frac{v^j_{nl} v^i_{lm}}{E_{lm}}\Biggr )\Biggr ].
  \end{aligned}
\end{equation}
The shift current conductivity $\sigma^{\rm{shift}}_{ijk}$ is symmetric with respect to the exchange of $j$ and $k$. 
For instance, $\sigma^{\rm{shift}}_{yxy}~=~\sigma^{\rm{shift}}_{yyx}$.

For linearly polarized light ($j=k$), this simplifies to
\begin{equation}
  \sigma_{ijj}^{\mathrm{shift}}
  = -\frac{i\hbar e^3}{S}\sum_{\bm{k}}\sum_{n,m}
    \frac{f_{nm}\,|v^j_{nm}|^2\,R^{ij}_{nm}}
         {E_{mn}^2(E_{mn}-\hbar\omega-i\eta)}.
\end{equation}
This expression clearly shows that the shift current is governed by the product of the transition intensity 
$\alpha_{mn}^{jj}=|v^j_{nm}|^2$ and the shift vector $R^{ij}_{nm}(\bm{k})$, 
which represents the real-space displacement of the electron wave packet during optical excitation. 
This geometric nature underlies the bulk photovoltaic effect~\cite{Morimoto2016_PRB}.
\end{widetext}

\subsection{Electronic Polarization and Berry Connection}
\label{app:polarization}
The electronic polarization can be expressed in terms of the Berry connection as
\begin{equation}
  P_i = e\int_{\rm{BZ}} \frac{d\bm{k}}{(2\pi)^3}\sum_m f_m \xi^i_{mm},
\end{equation}
where the Berry connection represents the Wannier center of wavefunctions in real space. 
This means that the real-space electric polarization is determined by the momentum-space Berry connection.
From this perspective, shift current originates from the difference in the charge center between the valence and conduction bands upon optical excitation.
However, shift current does not generally appear in all systems: it vanishes in centrosymmetric systems.
Even in noncentrosymmetric systems, the allowed tensor components of $\sigma_{ijk}^{\rm{shift}}(0;\omega, -\omega)$ are restricted by the crystal symmetry.

\section{Contour plots of shift vector and transition intensity}\label{app:shift_vector}
By constructing the effective model, 
Here, we show contour plots of the shift vector $R^{ij}_{nm}(\bm{k})$
and the transition intensity $\alpha^{jk}_{mn} = v^j_{mn} v^k_{nm}$.
This facilitates an intuitive understanding of the shift current conductivity $\sigma^{\rm{shift}}_{ijk}$,
analogous to the relationship between the Berry curvature and the transverse conductivity in linear optical responses.

\begin{figure*}
  \begin{center}
    \includegraphics[width=0.95\textwidth]{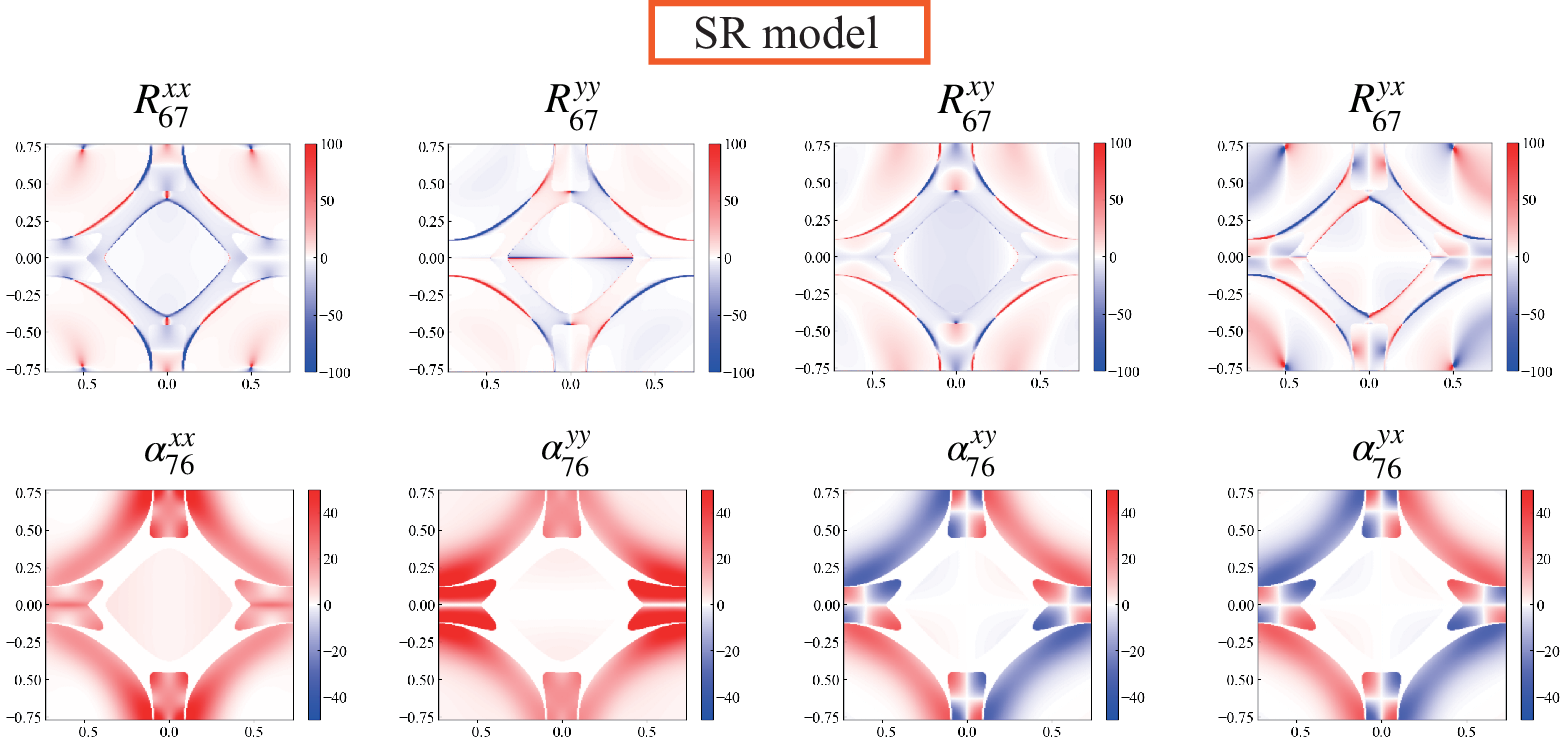}
    \caption{Contour plot of shift vector $R^{ij}_{nm}(\bm{k})$ and transition intensity 
    $\alpha^{jk}_{mn}=v^j_{mn}v^k_{nm}$ in the SR model.}
    \label{fig:R+alpha_SR}
  \end{center}
\end{figure*}

\begin{figure*}
  \begin{center}
    \includegraphics[width=0.95\textwidth]{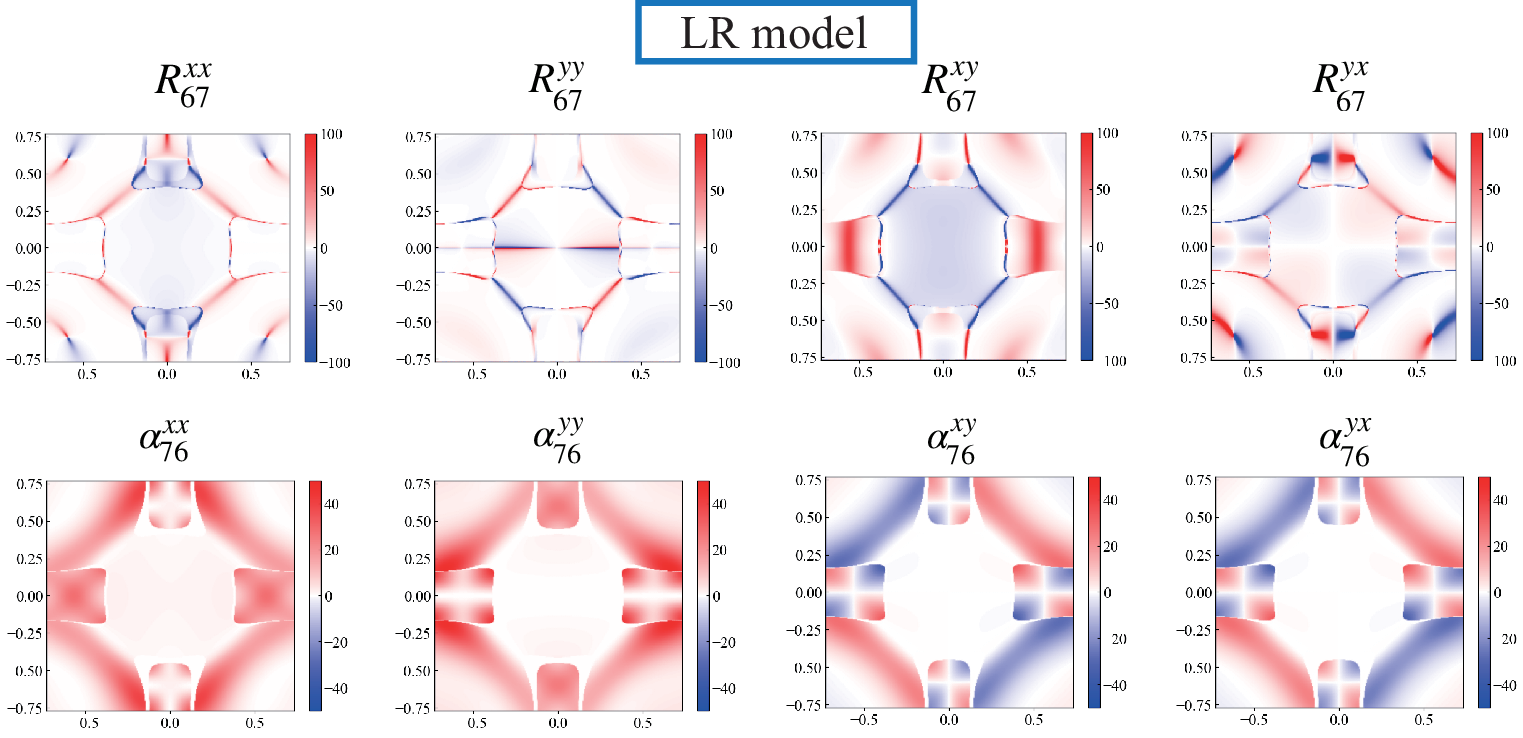}
    \caption{Contour plot of shift vector $R^{ij}_{nm}(\bm{k})$ and transition intensity 
    $\alpha^{jk}_{mn}=v^j_{mn}v^k_{nm}$ in the LR model.}
    \label{fig:R+alpha_LR}
  \end{center}
\end{figure*}

Figures~\ref{fig:R+alpha_SR} and \ref{fig:R+alpha_LR} show contour plots of 
$R^{ij}_{nm}(\bm{k})$ and $\alpha^{jk}_{mn}(\bm{k})=v^j_{mn}v^k_{nm}$ for $n=6, m=7$ in the SR and LR models, respectively. 
In Eq.~(\ref{eq:shift_current}), the non-integral function is expressed as the product
\begin{equation}
\alpha^{jk}_{mn}(\bm{k})R^{ik}_{nm}(\bm{k}) + \alpha^{kj}_{mn}(\bm{k})R^{ij}_{nm}(\bm{k}).
\end{equation}
This function determines whether $\sigma_{ijk}^{\rm{shift}}$ is finite. 
When it is an even function with respect to $k_x = 0$, $\sigma_{ijk}^{\rm{shift}}$ is finite, 
whereas when it is an odd function, $\sigma_{ijk}^{\rm{shift}}$
vanishes.

For $j=k$, the function simplifies to $R^{ij}_{nm}(\bm{k})\alpha^{jj}_{mn}(\bm{k})$. 
For example, in Fig.~\ref{fig:R+alpha_SR}, $R^{xx}_{67}(\bm{k})$ and $\alpha^{xx}_{76}(\bm{k})$ are both even functions, 
ensuring that $\sigma_{xxx}^{\rm{shift}}$ is finite. Similarly, $R^{xy}_{67}(\bm{k})$ and $\alpha^{yy}_{76}(\bm{k})$ are both even functions, 
confirming that $\sigma_{xyy}^{\rm{shift}}$ is finite, whereas for $R^{yy}_{67}(\bm{k})\alpha^{yy}_{76}(\bm{k})$, 
the product of an odd and even function is an odd function, leading to $\sigma_{yyy}^{\rm{shift}} = 0$. 
For $\sigma_{yyx}^{\rm{shift}}$, 
we examine the function $\alpha^{yx}_{76}(\bm{k})R^{yx}_{67}(\bm{k}) + \alpha^{xy}_{76}(\bm{k})R^{yy}_{67}(\bm{k})$. 
Since both $\alpha^{yx}_{76}(\bm{k})$ and $R^{yx}_{67}(\bm{k})$ are odd functions, their product is even. 
Similarly, $\alpha^{xy}_{76}(\bm{k})$ and $R^{yy}_{67}(\bm{k})$ are even functions, making their product even. 
Since the sum of these two terms is nonzero, $\sigma_{yyx}^{\rm{shift}}$ is finite. 
The same analysis can be applied to other conductivity components, yielding similar results for 
the LR model as shown in Fig.~\ref{fig:R+alpha_LR}.

\bibliography{reference}
\end{document}